\documentclass[a4paper,aps,nofootinbib]{revtex4}

\usepackage{graphicx}
\usepackage{amssymb}   
\usepackage{bm}        
\usepackage{amsmath}

\def\beq{\begin{equation}}
\def\eeq{\end{equation}}

\def\beqn{\begin{eqnarray}}
\def\eeqn{\end{eqnarray}}

\begin{document}

\title{Massless and Massive Gauge-Invariant Fields
in the Theory of Relativistic Wave Equations }

\author{V. A. Pletyukhov}
\affiliation{Brest State University, Brest, Belarus}
\email{pletyukhov@yandex.ru}
\author{V. I. Strazhev}
\affiliation{Belarusian State University, Minsk, Belarus}
\email{str-vi@mail.ru}

\begin{abstract}
In this work consideration is given to massless and massive gauge-invariant
spin 0 and spin 1 fields (particles) within the scope of a theory of the
generalized relativistic wave equations with an extended set of the Lorentz
group representations. The results obtained may be useful as regards the
application  of a relativistic
wave-equation theory in modern field models.
\end{abstract}

\maketitle

\section{ Introduction  }

One of the most extensively used ways to describe fundamental particles and
fields is still a theory of relativistic wave equations (RWE), the
foundations of which have been laid by Dirac [1], Fierz and Pauli [2,3],
Bhabha [4,5], Harish-Chandra [6,7], Gel'fand and Yaglom [8], Fedorov [9,10].
This theory has been advanced proceeding from the assumption that a
relativistic-invariant description of both massive and massless particles
(fields) may always be reduced to a system of the first-order differential
equations with constant factors, in the matrix form being given as follows:
\begin{equation}
\label{eq1.1}
\left( {\gamma _\mu  \partial _\mu  + \gamma _0 } \right)\psi
\left( x \right) = 0 \quad \left( {\mu =1\div 4} \right).
\end{equation}
Here $\psi \left( x \right)$ is multicomponent wave function transformed in
terms of some reducible Lorentz group representation $T$, $\gamma _\mu $ and
$\gamma _0 $ are square matrices.

In the case when the matrix $\gamma _0 $ is nonsingular $\left( {\det \gamma
_0 \ne 0} \right)$, equation (\ref{eq1.1}) describing a massive particle may be
reduced to the following form by multiplication into $m \gamma _0 ^{ - 1}$ :
\begin{equation}
\label{eq1.2}
\left( {\gamma _\mu  \partial _\mu    +   m  I} \right)  \psi
\left( x \right)    =    0  ,
\end{equation}
where $m$ is a parameter associated with mass, $I$ is unity matrix.

A choice of the matrices $\gamma _\mu $ in equations (\ref{eq1.1}) and
(\ref{eq1.2})
 is limited
by the following requirements (e.g., see [8,9]):

\noindent
~~i) invariance of the equation with respect to the transformations of its own
Lorentz group;

\noindent
~ii) invariance with respect to reflections;

\noindent
iii) possibility for derivation of the equation from the variational principle.

Equations of the form (\ref{eq1.2}) meeting requirements i)--iii) are known as
relativistic wave equations (RWE); equations of the form (\ref{eq1.1}) with the same
requirements are known as generalized RWE [9].

From requirement i) and from the condition of theory's irreducibility with
respect to the Lorentz group it follows that the function $\psi $ must be
transformed by some set of linking irreducible Lorentz-group
representations, forming what is known as a scheme for linking. The
representations $\tau    \sim \left( {l_1  ,  l_2 } \right)$ and
${\tau }' \sim \left( {{l}'_1 ,{l}'_2 } \right)$ are referred to
as linking if ${l}'_1  = l_1 \pm \frac{1}{2}, \quad {l}'_2
= l_2 \pm \frac{1}{2}.$

Aside from a choice of the wave function $\psi $, in definition of different
spin and mass states possible for the particle described by equations (\ref{eq1.1}) and
(\ref{eq1.2}) the matrices $\gamma _4 $ and $\gamma _0 $ are of particular
importance. Properties of the matrix $\gamma _4 $ are discussed
comprehensively in [8]. A structure of the matrix $\gamma _0 $ is determined
in [5,9]. Specifically, requirement i) results in reducibility of $\gamma _0
$ to the diagonal form, the matrix being composed of independent scalar
blocks corresponding to the irreducible representations of $\tau  $. For
$\det \gamma _0 = 0$ some of these blocks are zero. As follows from
requirement ii), nonzero elements $a_\tau $ of the matrix $\gamma _0 $
satisfy the relation
\begin{equation}
\label{eq1.3}
a_\tau = a_{\dot {\tau }} ,
\end{equation}
where $\dot {\tau }$ is representation conjugate to $\tau  $ with respect to
the spatial reflection, i.e., if $\tau  \sim  \quad \left( {l_1 ,l_2 }
\right)$, we have $\dot {\tau } \sim \left( {l_2 ,l_1 } \right)$.
In case of the finite-dimensional representations requirement iii) also leads
to the relation of (\ref{eq1.3}).

A distinctive feature of most well-known RWE of the form (\ref{eq1.2}) (Dirac
equation for spin $\frac{1}{2}$, Duffin-Kemmer equations for spins 0 and 1,
Fierz-Pauli equation for spin $\frac{3}{2})$ is the fact that they involve a
set of the Lorentz group representations \textit{minimally necessary}
for framing of a theory of this spin.

Such an approach in the case of$\det \gamma _0 = 0$ results in equations for
zero-mass particles (e.g., Maxwell equations). Because of this, selection of
$\det \gamma _0 = 0$ (also including $\gamma _0 = 0)$ in a theory of RWE is
associated with a description of massless particles [9,11].

It is known that, as distinct from the description of massive particles, in
a theory of massless particle with integer spin some of the wave-function
components are unobservable (potentials) and others - observable
(intensities). In consequence, for the potentials one can define the gauge
transformations and impose additional requirements excluding ``superfluous''
components of $\psi $. But for the description of massive particles by RWE
reducible to the form given by (\ref{eq1.2}),
the above-mentioned differentiation of
the wave-function components is not the case. In other words, the notion of
the gauge invariance of RWE (\ref{eq1.1}) in the sense indicated previously is
usually used for massless theories.

At the same time, there are papers, where so-called
\textit{massive gauge-invariant theories} are considered taking
other approaches. Illustrative examples are furnished by St\"uckelberg's
approach to the description of a massive spin 1 particle (see [12] and
references herein) and by a
$\mathord{\buildrel{\lower3pt\hbox{$\scriptscriptstyle\frown$}}\over {B}}
\wedge \mathord{\buildrel{\lower3pt\hbox{$\scriptscriptstyle\frown$}}\over
{F}} $-theory [13--16] claiming for the description of string interactions
in 4-dimensional space and suggesting a mechanism
(differing from Higgs's) of the mass
generation due to gauge-invariant mixing of electromagnetic and massless
vector fields with zero helicity. In the literature this field is called the
Kalb-Ramond field [15,16] and the notoph [17]. Because of this, one should
clear the question concerning the status of massive gauge-invariant fields
in the theory of RWE.

Another feature of well-known RWE is the fact that on going from equation of
the form (\ref{eq1.2}) for a massive spin $S$ particle
to its massless analog of the
form (\ref{eq1.1}), by making the substitution $mI \to \gamma _0 ,\det \gamma _0
= 0$, not all of the helicity values from $ + S$ to $ - S$ are retained, a
part of them is lost. This is the case when passing from the Duffin-Kemmer
equation for spin 1 to Maxwell equations with the dropped-out zero helicity.
In some modern models there is a necessity for simultaneous description of
different massless fields [18]. Within the scope of a theory of RWE, it
seems possible to solve this problem by the development of a scheme for
passage from (\ref{eq1.2}) to (\ref{eq1.1})
RWE with the singular matrix $\gamma _0$
retaining not only maximal but also intermediate helicity values.

By authors' opinion, solution of the stated problems is important
considering a possibility of using the well-developed apparatus of a theory
based on RWE in modern theoretical field models including the
phenomenological description of strings and superstrings in a space of the
dimension $d = 4$.

\section{GAUGE-INVARIANT THEORIES FOR MASSIVE SPIN 0 AND 1 PARTICLES}

Let us consider the following set of the Lorentz group irreducible
representations in a space of the wave function$\psi $
\begin{equation}
\label{eq2.1}
\left( {0,0} \right)\oplus \left( {\frac{1}{2},\frac{1}{2}}
\right) \oplus \left( {0,1} \right) \oplus \left( {1,0}
\right).
\end{equation}

The most general form of the corresponding (\ref{eq2.1}) tensor system of the
first-order equations meeting the requirements i) -- iii) is given by
\begin{subequations}
\label{eq2.2}
\beqn
\label{eq5}
\alpha  \partial _\mu  \psi _\mu + a\psi _0 =0,
\\
\label{eq6}
\beta ^ * \partial _\nu \psi _{\mu \nu } + \alpha^*
\partial _\mu \psi _0 + b\psi _\mu  = 0,
\\
\label{eq7}
\beta \left( { - \partial _\mu  \psi _\nu  + \partial _\nu  \psi
_\mu } \right) + c\psi _{\mu\nu }  = 0.
\eeqn
\end{subequations}
Here $\psi _0 $ is scalar, $\psi _\mu $ is vector, $\psi _{\mu \nu } $ is
antisymmetric second-rank tensor; $\alpha , \beta $ are arbitrary
dimensionless, generally speaking, complex parameters, and
$a,b,c$ are real nonnegative parameters, the dimension of which
on selection of $\hbar = c = 1$ is coincident with that of mass
(massive parameters). Writing system (\ref{eq2.2}) in the matrix form (\ref{eq1.1}), we
obtain in the basis
\begin{equation}
\label{eq2.3}
\psi  = \left( {\psi _0,\psi _\mu ,\psi _{\mu \nu
} } \right)   - column
\end{equation}
for the matrix $\gamma _0 $ the following expression:
\begin{equation}
\label{eq2.4}
\gamma _0  = \left( {{\begin{array}{*{20}c}
 a \hfill & \hfill & \hfill \\
 \hfill & {b I_4 } \hfill & \hfill \\
 \hfill & \hfill & {c I_6 } \hfill \\
\end{array} }} \right).
\end{equation}
(Matrices of the form $\gamma _\mu $ are not given as they are of no use for
us in further consideration.)

In the general case, when none of the parameters in (\ref{eq2.2}) is zero, this
system describes a particle with a set of spins 0, 1 and with two masses
\begin{equation}
\label{eq2.5}
m_1  = \frac{\sqrt {a b} }{\left| \alpha
\right|},\quad m_2  = \frac{\sqrt {b c} }{\left| \beta \right|},
\end{equation}
the mass $m_1 $ being associated with spin 0 and $m_2 $ with spin 1.
Omitting cumbersome calculations, we will verify this during analysis of
special cases.

Imposing on the parameters of system (\ref{eq2.2}) the requirement
\begin{equation}
\label{eq2.6}
\frac{\sqrt a }{\left| \alpha \right|} = \frac{\sqrt c }{\left|
\beta \right|},
\end{equation}
we obtain RWE for a particle with spins 0, 1 and one mass $m = m_1 = m_2 $.
At $\alpha  = 0$ system (\ref{eq2.2}) goes to the Duffin-Kemmer equation of
a particle with spin 1 and mass $m = m_2 $
\begin{subequations}
\label{eq2.7}
\beqn
\label{eq12}
\beta ^ * \partial _\nu \psi _{\mu  \nu }  + b\psi _\mu
= 0,
\\
\label{eq13_1}
\beta \left( { - \partial _\mu \psi _\nu  + \partial _\nu \psi
_\mu } \right) + c\psi _{\mu \nu }  = 0.
\eeqn
\end{subequations}

Finally, by setting in (\ref{eq2.2}) $\beta = 0$, we arrive at the
Duffin-Kemmer equation for a particle with spin 0 and mass $m=m_1$:
\begin{subequations}
\label{eq2.8}
\beqn
\label{eq13}
\alpha \partial _\mu \psi _\mu  + a\psi _0  =0,
\\
\label{eq14}
\alpha^* \partial _\mu \psi _0  + b\psi _\mu  =0.
\eeqn
\end{subequations}

Now we consider the case that is of great interest for us, when the
parameters $a,b,c$ determining a structure of the matrix $\gamma _0 $
in (\ref{eq2.4}) are selectively set to zero.

In system (\ref{eq2.2}) setting
\begin{equation}
\label{eq2.9}
a = 0,
\end{equation}
we have the following system of equations:
\begin{subequations}
\label{eq2.10}
\beqn
\label{eq17}
\partial _\mu  \psi _\mu  = 0,
\\
\label{eq18}
\beta ^ * \partial _\nu \psi _{\mu \nu }  + \alpha ^ *
\partial _\mu \psi _0  + b\psi _\mu  = 0,
\\
\label{eq19}
\beta \left( { - \partial _\mu\psi _\nu  + \partial _\nu \psi
_\mu } \right) + c\psi _{\mu \nu }  = 0,
\eeqn
\end{subequations}
that, being written in the matrix form of (\ref{eq1.1}),
corresponds in basis (\ref{eq2.3})
to the singular matrix $\gamma _0 $
\begin{equation}
\label{eq2.11}
\gamma _0      =     \left( {{\begin{array}{*{20}c}
 0 \hfill & \hfill & \hfill \\
 \hfill & {b I_4 } \hfill & \hfill \\
 \hfill & \hfill & {c I_6 } \hfill \\
\end{array} }} \right)    .
\end{equation}
From system (\ref{eq2.10}) one can easily derive the second-order equations
\beqn
\label{eq2.12}
\Box\psi _0 = 0
\\
\label{eq2.13}
\Box\psi _\mu  - \frac{c\alpha ^ * }{\left| \beta
\right|^2}\partial _{\mu } \psi _0 - \frac{b c}{\left| \beta
\right|^2}\psi _\mu  = 0.
\eeqn

As regards the scalar function $\psi _0 $ governed by equation (\ref{eq2.12}), the
following aspects must be taken into account. System (\ref{eq2.10}) is invariant
with respect to the gauge transformations
\begin{equation}
\label{eq2.14}
\psi _0  \to \psi _0  - \frac{1}{\alpha ^ * }\Lambda
,\quad \psi _\mu  \to \psi _\mu  +
\frac{1}{b}\partial _\mu \Lambda ,
\end{equation}
where the gauge function $\Lambda $ is limited by the constraint
\beq
\label{eq2.15}
\Box\Lambda=0.
\eeq
From comparison between (\ref{eq2.15}) and (\ref{eq2.12}) it follows that the function $\psi
_0 $ acts as a gauge function and hence provides no description for a
physical field. In other words, gauge transformations (\ref{eq2.14}) and
(\ref{eq2.15}) make it
possible to impose an additional condition
\begin{equation}
\label{eq2.16}
\psi _0  = 0.
\end{equation}
In this case system (\ref{eq2.10}) is transformed to system (\ref{eq2.7}) describing a
massive spin 1 particle, whereas equation (\ref{eq2.13}), considered simultaneously
with (\ref{eq17}), goes to an ordinary Proca equation. In this way the gauge
invariance of system (\ref{eq2.10}), as compared to (\ref{eq2.2}), leads to a decrease in
physical degrees of freedom from four to three, exclusive of the spin 0
state.

Note that a similar result may be obtained without the explicit use of the
considerations associated with the gauge invariance. By the introduction of
\begin{equation}
\label{eq2.17}
\varphi _\mu  = \psi _\mu  + \frac{\alpha ^ *
}{b}\partial _\mu \psi _0
\end{equation}
system (\ref{eq2.10}) may be directly reduced to the form
\begin{subequations}
\label{eq2.18}
\beqn
\label{eq24}
\beta ^ * \partial _\nu \psi _{\mu  \nu }  + b\varphi _\mu = 0,
\\
\label{eq25}
\beta \left( { - \partial _\mu\varphi _\nu  + \partial _\nu
\varphi _\mu } \right) + c\psi _{\mu \nu }  =0
\eeqn
\end{subequations}
coincident with (\ref{eq2.7}).

This variant of a gauge-invariant theory is known [12] as a Stueckelberg
approach to the description of a massive spin 1 particle. We have considered
this variant for a complete study of the possibilities given by system
(\ref{eq2.2}).

In (\ref{eq2.2}) we set
\begin{equation}
\label{eq2.19}
c= 0.
\end{equation}
Then the initial system of equations (\ref{eq2.2}) takes the form
\begin{subequations}
\label{eq2.20}
\beqn
\label{eq27}
\alpha \partial _\mu \psi _\mu  + a\psi _0  =0,
\\
\label{eq28}
\beta ^ * \partial _\nu \psi _{\mu \nu } + \alpha ^ *
\partial _\mu\psi _0  + b\psi _\mu = 0,
\\
\label{eq29}
 - \partial _\mu \psi _\nu  + \partial _\nu \psi _\mu =0.
\eeqn
\end{subequations}

According to (\ref{eq2.5}), it should describe a particle with the mass $m_1=
\frac{\sqrt {ab} }{\left| \alpha \right|}$ and with spin 0. By
convolution of equation (\ref{eq28}) with the operator $\partial _\mu $ we have
\beq
\label{eq2.21}
\Box\psi _0  + \frac{b}{\alpha ^ * }\partial _\mu \psi _\mu = 0.
\eeq
Comparing (\ref{eq2.21}) with (\ref{eq27}), we arrive at the equation
\beq
\label{eq2.22}
\psi _0  - \frac{ab}{\left| \alpha \right|^2}\psi _0  =0,
\eeq
that provides support for all the afore-said.

The states associated with spin 1, for the condition set by (\ref{eq2.19}),
disappear
due to the invariance of system (\ref{eq2.20}) with respect to the gauge
transformations
\begin{equation}
\label{eq2.23}
\psi _{\mu \nu }  \to \psi _{\mu \nu }  -
\frac{1}{\beta ^ * }\Lambda _{\mu \nu },\quad\psi _\mu
 \to \psi _\mu + \frac{1}{b}\partial _\nu \Lambda
_{\mu \nu },
\end{equation}
where an arbitrary choice of the gauge function $\Lambda _{\mu \nu } $
is constrained by
\begin{equation}
\label{eq2.24}
\partial _\alpha  \partial _\nu  \Lambda _{\mu  \nu }  - \partial
_\mu  \partial _\nu  \Lambda _{\alpha  \nu } = 0.
\end{equation}
On the other hand, as follows from equations (\ref{eq28}), (\ref{eq29}),
a similar equation
\begin{equation}
\label{eq2.25}
\partial _\alpha \partial _\nu  \psi _{\mu \nu }  - \partial
_\mu \partial _\nu \psi _{\alpha \nu }  = 0
\end{equation}
is satisfied by the tensor $\psi _{\mu  \nu } $\textbf{.} Consequently, a
choice of $\Lambda _{\mu \nu } $ is arbitrary enough to impose an
additional constraint
\begin{equation}
\label{eq2.26}
\partial _\nu \psi _{\mu  \nu }  = 0
\end{equation}
that is in accord with (\ref{eq2.25}). In this case system (\ref{eq2.20})
takes the form of
(\ref{eq2.8}), i.e. it actually describes a massive spin 0 particle.

Note also that system (\ref{eq2.20}) may be reduced to the form
\begin{subequations}
\label{eq2.27}
\beqn
\label{eq34}
\alpha \partial _\mu \varphi _\mu  + a\psi _0 =0,
\\
\label{eq35}
\alpha ^{_ * }\partial _\mu \psi _0 + b\varphi _\mu  =0
\eeqn
\end{subequations}
similar to (\ref{eq2.8}) by introduction of the vector
\begin{equation}
\label{eq2.28}
\varphi _\mu  = \psi _\mu  + \frac{\beta ^ *
}{b}\partial _\nu \psi _{\mu \nu }.
\end{equation}

Thus, the considered variant of a massive gauge-invariant theory is some
kind of an analog for the St\"uckelberg approach but applicable to the
description of a spin 0 particle. The authors have not found any mentioning
of such a description in the literature available.

In the formalism of RWE (\ref{eq1.1}) this theory is consistent with the matrix
$\gamma _0 $ of the form
\begin{equation}
\label{eq2.29}
\gamma _0  = \left( {{\begin{array}{*{20}c}
 a \hfill & \hfill & \hfill \\
 \hfill & {b I_4 } \hfill & \hfill \\
 \hfill & \hfill & {0_6 } \hfill \\
\end{array} }} \right).
\end{equation}

Next we consider a set of the Lorentz group representations
\begin{equation}
\label{eq2.30}
\left( {\frac{1}{2}, \frac{1}{2}} \right)   \oplus   \left(
{\frac{1}{2}, \frac{1}{2}} \right)^\prime    \oplus   \left( {0, 1}
\right)   \oplus   \left( {1, 0} \right)  ,
\end{equation}
where the representation $\left( {\frac{1}{2} ,  \frac{1}{2}}
\right)^\prime $conforms to the pseudovector or to the absolutely
antisymmetric third-rank tensor. The most general form of a tensor system of
the first-order equations based on representation (\ref{eq2.30}) and meeting the
above-mentioned requirements i)--iii) is given by
\begin{subequations}
\label{eq2.31}
\beqn
\label{eq39}
\alpha  \partial _\nu  \psi _{\mu   \nu }    +   a \psi _\mu
=    0  ,
\\
\label{eq40}
\beta  \partial _\nu  \tilde {\psi }_{\mu   \nu }    +   b \tilde
{\psi }_\mu     =    0  ,
\\
\label{eq41}
\alpha ^ *  \left( { -  \partial _\mu  \psi _\nu    +   \partial _\nu
 \psi _\mu } \right)   +   \beta ^ *  \varepsilon _{\mu  \nu  \alpha
 \beta }  \partial _\alpha  \tilde {\psi }_\beta    +   c \psi _{\mu
 \nu }     =    0  .
\eeqn
\end{subequations}
Here $\tilde {\psi }_{\mu  \nu }    =   \frac{1}{2} \varepsilon _{\mu
 \nu  \alpha  \beta }  \psi _{\alpha  \beta }  , \quad \tilde {\psi }_{\mu
 }    =   \frac{1}{6} \varepsilon _{\mu  \nu  \alpha  \beta }
 \psi _{\nu  \alpha  \beta }  , \quad  \varepsilon _{\mu  \nu  \alpha
 \beta }   $ is the Levi-Civita tensor ($\varepsilon _{1 2 3 4}  = -
 i)$, $\psi _{\nu  \alpha \beta } $ is antisymmetric third-rank tensor,
$\alpha , \beta $ are still arbitrary dimensionless, generally speaking,
complex parameters, $a,b,c$ are mass parameters.

Writing system (\ref{eq2.31}) in the form (\ref{eq1.1}), where $\Psi = \left( {\psi _\mu
,  \tilde {\psi }_\mu , \psi _{\mu  \nu } } \right)$ is column, for the
matrix $\gamma _0 $ we get the expression
\begin{equation}
\label{eq2.32}
\gamma _0     =    \left( {{\begin{array}{*{20}c}
 {a I_4 } \hfill & \hfill & \hfill \\
 \hfill & {b I_4 } \hfill & \hfill \\
 \hfill & \hfill & {c I_6 } \hfill \\
\end{array} }} \right).
\end{equation}

Now we elaborate on massive gauge-invariant theories obtainable from (\ref{eq2.31})
by manipulations with the parameters $a, b, c$.

Let us take the case
\begin{equation}
\label{eq2.33}
a    =    0 .
\end{equation}
In this case we have a system of equations
\begin{subequations}
\label{eq2.34}
\beqn
\label{eq44}
 \partial _\nu  \psi _{\mu   \nu }       =    0  ,
\\
\label{eq45}
\beta  \partial _\nu  \tilde {\psi }_{\mu   \nu }    +   b \tilde
{\psi }_\mu     =    0  ,
\\
\label{eq46}
\alpha ^ *  \left( { -  \partial _\mu  \psi _\nu    +   \partial _\nu
 \psi _\mu } \right)   +   \beta ^ *  \varepsilon _{\mu  \nu  \alpha
 \beta }  \partial _\alpha  \tilde {\psi }_\beta    +   c \psi _{\mu
  \nu }     =    0
\eeqn
\end{subequations}
that, when formulated as (\ref{eq1.1}), is associated with the singular
matrix $\gamma _0 $
\begin{equation}
\label{eq2.35}
\gamma _0     =    \left( {{\begin{array}{*{20}c}
 {0_4 } \hfill & \hfill & \hfill \\
 \hfill & {b I_4 } \hfill & \hfill \\
 \hfill & \hfill & {c I_6 } \hfill \\
\end{array} }} \right)   .
\end{equation}

From (\ref{eq2.34}) we can obtain the second-order equations
\beqn
\label{eq2.36}
\Big( \Box - \frac{b c}{|\beta|^2}\Big)\tilde {\psi }_\mu =0,
\\
\label{eq2.37}
\partial _\mu  \tilde {\psi }_\mu     =    0  ,
\\
\label{eq2.38}
\Box\psi _\mu    -   \partial _\mu  \partial _\nu  \psi _\nu     =
   0  .
\eeqn
Equations (\ref{eq2.36}) and (\ref{eq2.37}) denote that system (\ref{eq2.34}) involves the description
of a massive spin 1 particle. As shown by equation (\ref{eq2.38}), system (\ref{eq2.34})
describes also a massless field with the potential $\psi _\mu $. The latter
allows for involvement of the gauge transformation
\begin{equation}
\label{eq2.39}
\psi _\mu     \to    \psi _\mu    +   \partial _\mu  \Lambda
\end{equation}
($\Lambda $ is arbitrary function), with respect to which system (\ref{eq2.34}) and
equation (\ref{eq2.38}) are invariant. The indicated invariance means that this
massless field is a Maxwell-type field with helicity $\pm 1$.

In this manner the gauge-invariant system (\ref{eq2.34}) irreducible with respect to
the Lorentz group offers a simultaneous description of a massive spin 1
particle and of a massless field with helicity $\pm 1$. In other words, here
we deal with a massive-massless gauge-invariant theory rather than massive
theory, as is the case for (\ref{eq2.10}) and (\ref{eq2.20}).

A similar result may be obtained if we set in (\ref{eq2.34})
\begin{equation}
\label{eq2.40}
b    =    0 .
\end{equation}
Then we have
\begin{equation}
\label{eqeq2.41}
\gamma _0     =    \left( {{\begin{array}{*{20}c}
 {a I_4 } \hfill & \hfill & \hfill \\
 \hfill & {0_4 } \hfill & \hfill \\
 \hfill & \hfill & {c I_6 } \hfill \\
\end{array} }} \right)   ,
\end{equation}
and the second-order equations following from the corresponding first-order
system
\begin{subequations}
\label{eq2.42}
\beqn
\label{eq52}
\alpha  \partial _\nu  \psi _{\mu   \nu }    +   a \psi _\mu
=    0  ,
\\
\label{eq53}
 \partial _\nu  \tilde {\psi }_{\mu   \nu }     =    0  ,
\\
\label{eq54}
\alpha ^ *  \left( { -  \partial _\mu  \psi _\nu    +   \partial _\nu
 \psi _\mu } \right)   +   \beta ^ *  \varepsilon _{\mu  \nu  \alpha
 \beta }  \partial _\alpha  \tilde {\psi }_\beta    +   c \psi _{\mu
 \nu }     =    0
\eeqn
\end{subequations}
are of the form
\beqn
\label{eq2.43}
\Big(\Box - \frac{a c}{\left| \alpha \right|^2}\Big)\psi _\mu =0,
\\
\label{eq2.44}
\partial _\mu  \psi _\mu     =    0  ,
\\
\label{eq2.45}
\Box\tilde {\psi }_\mu    -   \partial _\mu  \partial _\nu  \tilde {\psi
}_\nu     =    0  .
\eeqn
Equation (\ref{eq2.45}) and system (\ref{eq2.42}) are invariant with respect to the gauge
transformations
\begin{equation}
\label{eq2.46}
\tilde {\psi }_\mu     \to    \tilde {\psi }_\mu    +   \partial
_\mu  \tilde {\Lambda }  .
\end{equation}
Thus, here we deal again with a gauge-invariant massive-massless spin 1
theory.

Let us consider another set of representations
\begin{equation}
\label{eq2.47}
\left( {0, 0} \right)^\prime    \oplus   \left(
{\frac{1}{2}, \frac{1}{2}} \right)^\prime    \oplus   \left( {0, 1}
\right)   \oplus   \left( {1, 0} \right) \quad ,
\end{equation}
where $\left( {0, 0} \right)^\prime $ is associated with the absolutely
antisymmetric fourth-rank tensor $\psi _{\mu  \nu  \alpha  \beta } $. The
most general tensor formulation of RWE based on the set of representations
given in (\ref{eq2.47}) takes the form
\begin{subequations}
\label{eq2.48}
\beqn
\label{eq58}
\alpha  \partial _{\left[ \mu \right.}  \psi _{\nu  \alpha  \beta \left.
\right]}    +   a \psi _{\mu  \nu  \alpha  \beta }     =
   0  ,
\\
\label{eq59}
\alpha ^ *  \partial  _{\left[ \nu \right.}  \psi _{\alpha  \beta \left.
\right]}     +    \beta ^ *  \partial _\mu  \psi _{\mu  \nu
 \alpha  \beta }    +   b \psi _{\nu  \alpha  \beta }     =
   0  ,
\\
\label{eq60}
\beta  \partial _\nu  \psi _{\nu  \alpha  \beta }     +    c \psi
_{\alpha  \beta }     =    0  ,
\eeqn
\end{subequations}
where the following notation is used:
\beqn
\label{eq2.49}
\partial _{\left[ \nu \right.}  \psi _{\alpha  \beta \left. \right]}
    \equiv    \partial _\nu  \psi _{\alpha  \beta }    +
  \partial _\beta  \psi _{\nu  \alpha }    +   \partial _\alpha
 \psi _{\beta  \nu }   ,
\\
\label{eq62}
\partial _{\left[ \mu \right. } \psi _{\nu  \alpha  \beta \left. \right]}
   \equiv   \partial _\mu  \psi _{\nu  \alpha  \beta }    -
  \partial _\nu  \psi _{\mu  \alpha  \beta }    +   \partial
_\alpha  \psi _{\mu  \nu  \beta }    -   \partial _\beta  \psi _{\mu
 \nu  \alpha }   .
\eeqn

After introduction into system (\ref{eq2.48}) of the dual conjugates $\tilde
{\psi }_{\mu  \nu }  ,   \tilde {\psi }_\mu $ and pseudoscalar $\tilde
{\psi }_0    =   \frac{1}{4\mbox{!}} \varepsilon _{\mu  \nu  \alpha
 \beta }  \psi _{\mu  \nu  \alpha  \beta } $ instead of the tensors $\psi
_{\mu  \nu }  ,   \psi _{\nu  \alpha  \beta  } ,   \psi _{\mu
 \nu  \alpha  \beta }  $, it is conveniently rewritten to give
\begin{subequations}
\label{eq2.50}
\beqn
\label{eq63}
\alpha  \partial _\mu  \tilde {\psi }_{\mu  }    +   a \tilde {\psi
}_0     =    0  ,
\\
\label{eq64}
 \beta ^ *  \partial _\nu  \tilde {\psi }_{\mu   \nu }    +
  \alpha ^ *  \partial _\mu  \tilde {\psi }_0    +   b \tilde {\psi
}_\mu    =    0  ,
\\
\label{eq65}
\beta   \left( { -  \partial _\mu  \tilde {\psi }_\nu    +
  \partial _\nu  \tilde {\psi }_\mu } \right)     +    c \tilde
{\psi }_{\mu  \nu }     =    0  .
\eeqn
\end{subequations}
As seen from the comparison between (\ref{eq2.50}) and
 (\ref{eq2.2}), these systems are dual
in that one may be derived from the other by the substitutions
\begin{equation}
\label{eq2.51}
\psi _0     \leftrightarrow    \tilde {\psi }_0  ,      \psi
_\mu     \leftrightarrow    \tilde {\psi }_\mu  ,      \psi
_{\mu  \nu }     \leftrightarrow    \tilde {\psi }_{\mu  \nu } .
\end{equation}

Clearly, the use of system (\ref{eq2.50}) with the aim of framing various
gauge-invariant theories on its basis follows the same procedure and gives
the same results as with system (\ref{eq2.2}). So, when in (\ref{eq2.50}) we set $a=0$,
a gauge-invariant theory for a pseudoscalar particle of the
mass $\frac{\sqrt {b c} }{\left| \beta \right|}$ is put forward. But
setting $c    =    0$, we arrive at a gauge-invariant theory for a
pseudoscalar particle of the mass $\frac{\sqrt {a b} }{\left| \alpha
\right|}$.

\section{SIMULTANEOUS DESCRIPTION OF MASSLESS FIELDS}

Returning to a set of representations (\ref{eq2.1}) and to tensor system (\ref{eq2.2}), we
consider the case
\begin{equation}
\label{3.1}
b    =    0 .
\end{equation}
The following system is obtained:
\begin{subequations}
\label{3.2}
\beqn
\label{3.2a}
\alpha  \partial _\mu  \psi _\mu    +   a \psi _0     =
   0  ,
\\
\label{3.2b}
\beta ^ *  \partial _\nu  \psi _{\mu  \nu }    +   \alpha ^ *
 \partial _\mu  \psi _0     =    0  ,
\\
\label{3.2c}
\beta  \left( { - \partial _\mu  \psi _\nu    +   \partial _\nu  \psi
_\mu } \right)   +   c \psi _{\mu  \nu }     =    0
\eeqn
\end{subequations}
that in basis (\ref{eq2.3}) is associated with the matrix $\gamma _0 $ of the form
\begin{equation}
\label{3.3}
\gamma _0     =    \left( {{\begin{array}{*{20}c}
 a \hfill & \hfill & \hfill \\
 \hfill & {0_4 } \hfill & \hfill \\
 \hfill & \hfill & {c I_4 } \hfill \\
\end{array} }} \right)   .
\end{equation}
From system (\ref{3.2}) we get d'Alembert equation (\ref{eq2.12}) for the scalar function
$\psi _0 $ and the second-order equation
\beq
\label{3.4}
\Box\psi _\mu    -   \left( {1   -   \frac{c \left| \alpha
\right|^2}{a\left| \beta \right|^2}} \right)  \partial _\mu  \partial
_\nu  \psi _\nu = 0
\eeq
for the vector $\psi _\mu $. From this it is inferred that we deal with a
massless field. When considering the quantities $\psi _0 $ and $\psi _\mu $
as potentials of this field, we treat equation (\ref{3.2c})
as a definition of the
intensity $\psi _{\mu  \nu } $ in terms of the potentials, (\ref{3.2a}) is
additional constraint similar to the Feynman gauge. Then equation (\ref{3.2b})
acts as an equation of motion.

With this treatment, system (\ref{3.2}) and equation (\ref{3.4}) is invariant with
respect to the gauge transformation
\begin{equation}
\label{3.5}
\psi _\mu     \to    \psi _\mu    +   \partial _\mu  \Lambda ,
\end{equation}
where an arbitrary choice of $\Lambda $ is constrained by (\ref{eq2.15}). Gauge
transformations (\ref{3.5}) and (\ref{eq2.15}) in combination with an additional requirement
(\ref{3.2a}) indicate that, among the four components of the potential $\psi _\mu
$, only two components are independent. They describe a transverse component
of the field under study. One more, longitudinal, component of this field is
described by the scalar function $\psi _0 $. In this way a choice of (\ref{3.1}) in
system (\ref{eq2.2}) leads to a theory of a massless filed with three helicity
values $\pm  1 ,  0 .$ This is one of the distinguishing features of
system (\ref{eq2.2}) as opposed to a theory of Duffin--Kemmer for spin 1, that on
a similar passage to the limit results in a massless field with helicities
$\pm  1 .$

Also, note that equation (\ref{3.4}) with due regard for (\ref{3.2a}) may be rewritten
as
\beq
\label{3.6}
\Box\psi _\mu    +   \left( {1   -   \frac{c \left| \alpha
\right|^2}{a\left| \beta \right|^2}} \right)   \frac{a}{\alpha
}  \partial _\mu   \psi _0     =    0  ,
\eeq
from whence it follows that a gradient of the scalar component acts as an
(internal) source of the transverse component of this massless field.

Next we select the case when in system (\ref{eq2.2})
\begin{equation}
\label{3.7}
a    =    0 ,   \quad   b    =    0  .
\end{equation}
The resultant system
\begin{subequations}
\label{3.8}
\beqn
\label{3.8a}
 \partial _\mu  \psi _\mu       =    0  ,
\\
\label{3.8b}
\beta ^ *  \partial _\nu  \psi _{\mu  \nu }    +   \alpha ^ *
 \partial _\mu  \psi _0     =    0  ,
\\
\label{3.8c}
\beta  \left( { - \partial _\mu  \psi _\nu    +   \partial _\nu  \psi
_\mu } \right)   +   c \psi _{\mu  \nu }     =    0
\eeqn
\end{subequations}
is distinguished from system (\ref{3.2}) by the potential gauge requirement
(compare (\ref{3.2a}) with (\ref{3.8a})).
In this case the matrix $\gamma _0 $ is of the form
\begin{equation}
\label{3.9}
\gamma _0     =    \left( {{\begin{array}{*{20}c}
 0 \hfill & \hfill & \hfill \\
 \hfill & {0_4 } \hfill & \hfill \\
 \hfill & \hfill & {c I_6 } \hfill \\
\end{array} }} \right)   .
\end{equation}
From (\ref{3.8}) one can obtain equation (\ref{eq2.12}) for the function $\psi _0 $ and
the second-order equation
\beq
\label{3.10}
\Box\psi _\mu    -   \frac{\alpha ^ * c}{\left| \beta
\right|^2}   \partial _\mu   \psi _0     =    0
\eeq
for $\psi _\mu $that, similar to system (\ref{3.8}),
is invariant with respect to
gauge transformations (\ref{3.5}), (\ref{eq2.15}).
All this indicates that we deal again
with two interrelated massless fields: vector field with helicity $\pm
 1 $ and scalar field with helicity $  0 $, the gradient of a scalar
field acting as a source of the vector field.

The other two massless analogs of system (\ref{eq2.2}), when
\begin{equation}
\label{3.11}
a    =    0 , \quad     c    =    0
\end{equation}
and
\begin{equation}
\label{3.12}
b    =    0 ,  \quad    c    =    0  ,
\end{equation}
are associated with the description of a massless field of zero helicity.
Establishing this fact, we will not concern ourselves with the details.

Considering the possibility for simultaneous description of different
massless fields, we next analyze a set of the representations in (\ref{eq2.30}) and
the first-order system of (\ref{eq2.31}).

First, we take the case
\begin{equation}
\label{3.13}
c = 0,  \quad   a = b .
\end{equation}
In this case system (\ref{eq2.31}) is of the form
\begin{subequations}
\label{3.14}
\beqn
\label{3.14a}
\alpha  \partial _\nu  \psi _{\mu  \nu }    +   a \psi _\mu  =
0,
\\
\label{3.14b}
\beta  \partial _\nu  \tilde {\psi }_{\mu  \nu }    +   a  \tilde
{\psi }_\mu  = 0,
\\
\label{3.14c}
\alpha ^ *  \left( { - \partial _\mu  \psi _\nu    +   \partial _\nu
 \psi _\mu } \right)   +   \beta ^ *  \varepsilon _{\mu  \nu  \alpha
 \beta }  \partial _\alpha  \tilde {\psi }_\beta = 0,
\eeqn
\end{subequations}
and the matrix $\gamma _0 $ (\ref{eq2.32}) is transformed to the matrix
\begin{equation}
\label{3.15}
\gamma _0     =   \left( {{\begin{array}{*{20}c}
 {aI_8 } \hfill & \hfill \\
 \hfill & {O_6 } \hfill \\
\end{array} }} \right).
\end{equation}
In (\ref{3.14}) we take components of the tensor $\psi _{\mu  \nu } $ as
potentials, assuming the vector $\psi _\mu $ and the pseudovector $\tilde
{\psi }_\mu $ as intensities. Then equations (\ref{3.14a}) and (\ref{3.14b})
 are the
intensity definitions in terms of the potentials and (\ref{3.14c}) acts as an
equation of motion.

From system (\ref{3.14}) we derive the second-order equation for the
tensor-potential $\psi _{\mu   \nu } $
\beq
\label{3.16}
\Box\psi _{\mu   \nu } = 0.
\eeq
Equations (\ref{3.14}) and (\ref{3.16}) are invariant with respect to the gauge
transformations
\begin{equation}
\label{3.17}
\psi _{\mu  \nu }     \to    \psi _{\mu  \nu }    +   \partial
_\mu  \Lambda _\nu    -   \partial _\nu  \Lambda _\mu   ,
\end{equation}
where an arbitrary choice of the functions $\Lambda _\mu $ is constrained by
\beq
\label{3.18}
\Box\Lambda _{\mu  }    -   \partial _\mu  \partial _\nu  \Lambda _\nu
    =    0  .
\eeq
Equation (\ref{3.16}) and gauge transformations (\ref{3.17}) and (\ref{3.18}) indicate that a
choice of (\ref{3.13}) leads to a theory for a massless particle of zero helicity
carrying spin 1 in the process of interactions.

By the present time, two approaches to the description of such a particle
have been known: (1) Ogievetsky and Polubarinov approach [17] in which an
intensity is represented by the vector (in [17] this particle is called the
notoph) and (2) Kalb-Ramond approach [13], where an intensity is represented
by the antisymmetric third-rank tensor or pseudovector (Kalb-Ramond field).
System (\ref{3.14}) combines the description of both fields in one irreducible
RWE.

In a sense this pattern may be complemented if in (\ref{eq2.31}) we set
\begin{equation}
\label{3.19}
a    =    0 ,  \quad    b    =    0  .
\end{equation}
As a result, we have the following system:
\begin{subequations}
\label{3.20}
\beqn
\label{3.20a}
\partial _\nu  \psi _{\mu  \nu }       =    0  ,
\\
\label{3.20b}
 \partial _\nu  \tilde {\psi }_{\mu  \nu }       =    0  ,
\\
\label{3.20c}
\alpha ^ *  \left( { - \partial _\mu  \psi _\nu    +   \partial _\nu
 \psi _\mu } \right)   +   \beta ^ *  \varepsilon _{\mu  \nu  \alpha
 \beta }  \partial _\alpha  \tilde {\psi }_\beta    +   c \psi _{\mu
 \nu }     =    0
\eeqn
\end{subequations}
that is associated with the matrix $\gamma _0 $ of the form
\begin{equation}
\label{3.21}
\gamma _0     =   \left( {{\begin{array}{*{20}c}
 {O_8 } \hfill & \hfill \\
 \hfill & {cI_6 } \hfill \\
\end{array} }} \right)    .
\end{equation}
In system (\ref{3.20}) the components $\psi _\mu $ and $\tilde {\psi }_\mu $ are
naturally considered as potentials, and $\psi _{\mu   \nu } $ is taken as
an intensity. Then it is invariant with the gauge transformations
\begin{equation}
\label{3.22}
\psi _\mu     \to    \psi _\mu    +   \Lambda _\mu
  ,        \tilde {\psi }_\mu     \to    \tilde {\psi }_\mu
   +   \tilde {\Lambda }_\mu   ,
\end{equation}
where an arbitrary choice of the gauge functions $\Lambda _\mu  ,  \tilde
{\Lambda }_\mu $ is constrained by
\begin{equation}
\label{3.23}
\alpha ^ *  \left( { - \partial _\mu  \Lambda _\nu    +   \partial
_\nu  \Lambda _\mu } \right)   +   \beta ^ *  \varepsilon _{\mu  \nu
 \alpha  \beta }  \partial _\alpha  \tilde {\Lambda }_\beta     =
   0  .
\end{equation}
In other words, at $\alpha    =   \beta    =   1$ system (\ref{3.20})
represents the well-known two-potential formulation from electrodynamics
(e.g., see [19]) for a massless spin 1 field with helicity $\pm $1.

Thus, a reciprocal complementarity of the theories based on systems (\ref{3.14})
and (\ref{3.20}) is exhibited in their mathematical structure, including that of
the matrix $\gamma _0 $, and also in interpretations of the field components
$\psi _\mu $, $\tilde {\psi }_\mu $, $\psi _{\mu   \nu } $ as well as in
properties (helicity) of the particles described.

Of particular interest is the case when in (\ref{eq2.31}) we set
\begin{equation}
\label{3.24}
a    =    0 ,\quad      c    =    0  .
\end{equation}
This results in the system
\begin{subequations}
\label{3.25}
\beqn
\label{3.25a}
\partial _\nu  \psi _{\mu  \nu }       =    0  ,
\\
\label{3.25b}
\beta   \partial _\nu  \tilde {\psi }_{\mu  \nu }    +   b \tilde
{\psi }_\mu     =    0  ,
\\
\label{3.25c}
\alpha ^ *  \left( { - \partial _\mu  \psi _\nu    +   \partial _\nu
 \psi _\mu } \right)   +   \beta ^ *  \varepsilon _{\mu  \nu  \alpha
 \beta }  \partial _\alpha  \tilde {\psi }_\beta       =
   0
\eeqn
\end{subequations}
and leads to the matrix
\begin{equation}
\label{3.26}
\gamma _0     =    \left( {{\begin{array}{*{20}c}
 {O_4 } \hfill & \hfill & \hfill \\
 \hfill & {b I_4 } \hfill & \hfill \\
 \hfill & \hfill & {O_6 } \hfill \\
\end{array} }} \right)   .
\end{equation}
For convenience, we rewrite (\ref{3.25}) in the following form:
\begin{subequations}
\label{3.27}
\beqn
\label{3.27a}
\partial _\nu  \psi _{\mu  \nu }       =    0  ,
\\
\label{3.27b}
\beta  \left( {\partial _\mu  \psi _{\nu  \alpha }    +   \partial
_\alpha  \psi _{\mu  \nu }    +   \partial _\nu  \psi _{\alpha  \mu
} } \right)   +   b \psi _{\mu  \nu  \alpha }     =    0  ,
\\
\label{3.27c}
\alpha ^ *  \left( { - \partial _\nu  \psi _\alpha    +   \partial
_\alpha  \psi _\nu } \right)   +   \beta ^ *  \partial _\mu  \psi
_{\mu  \nu  \alpha }       =    0  ,
\eeqn
\end{subequations}
where $\psi _{\mu   \nu  \alpha } $ is antisymmetric third-rank tensor
dual with respect to the pseudovector $\tilde {\psi }_\mu $.

According to the structure of system (\ref{3.27}), $\psi _\mu $ and $\psi _{\mu
  \nu } $ are potentials, $\psi _{\mu   \nu  \alpha } $ is intensity.
Then equation (\ref{3.27b}) is a definition of the intensity, and (\ref{3.27a})
 acts as
an additional constraint imposed on the tensor-potential $\psi _{\mu  \nu }
$ and included originally in the system itself. This constraint leaves for
tensor $\psi _{\mu  \nu } $ satisfying the second-order equation
\beq
\label{3.28}
\Box\psi _{\mu  \nu }    +    \frac{\left| \alpha \right|^2}{\left|
\beta \right|^2}  \frac{b}{\alpha } \left( {\partial _\mu  \psi _\nu
   -   \partial _\nu  \psi _\mu } \right)    =    0
\eeq
two independent components. As this takes place, system (\ref{3.27}) is invariant
with respect to relative gauge transformations (\ref{3.17}), (\ref{3.18}). Due to an
arbitrary choice of the gauge function $\Lambda _\mu  $ constraining by
condition (\ref{3.18}) we have only one independent component for $\psi _{\mu
 \nu } $ that is associated with the state of a massless field with zero
helicity.

To elucidate a meaning of the term $ \partial _\mu  \psi _\nu   -
 \partial _\nu  \psi _\mu    $ in (\ref{3.28}), we turn to the potential $\psi
_{\mu  } $. Apart from transformations (\ref{3.17}), (\ref{3.18}), system (\ref{3.27}) is
also invariant with respect to the gauge transformation
\begin{equation}
\label{3.29}
\psi _\mu     \to    \psi _\mu    +   \partial _\mu  \Lambda
  ,
\end{equation}
where $\Lambda  $ is arbitrary function . From equation (\ref{3.27c}) for $\psi
_{\mu  } $ we derive the second-order equation
\beq
\label{3.30}
\Box\psi _{\mu  }   -    \partial _\mu  \partial _\nu  \psi _\nu
      =    0,
\eeq
in combination with (\ref{3.18}) indicating that the potential $\psi _{\mu  } $
gives description for the transverse component (helicity $\pm 1$) of the
massless field under study. The expression
\begin{equation}
\label{3.31}
 \partial _\mu  \psi _\nu    -    \partial _\nu  \psi _\mu
\equiv    F_{\mu  \nu }
\end{equation}
in equations (\ref{3.27c}) and (\ref{3.28}) may be considered as an intensity associated
with this transverse component. Then equation (\ref{3.27c}) rewritten with regard
to the notation of (\ref{3.31}) as
\begin{equation}
\label{3.32}
\beta ^ *  \partial _\mu  \psi _{\mu \nu  \alpha }    -    \alpha ^
*  F_{\nu  \alpha }     =    0  ,
\end{equation}
acts as an equation of motion in system (\ref{3.27}).

Thus, a choice (\ref{3.24}) of mass parameters in the initial system (\ref{eq2.31}) leads
to a theory of the generalized massless field with polarizations 0, $\pm $1.

Selection of the parameters
\begin{equation}
\label{3.33}
b    =    0 ,   \quad   c    =    0 .
\end{equation}
in system (\ref{eq2.31}) also results in a theory of the generalized massless field
with helicities 0, $\pm $1 featuring a dual conjugate of that obtainable in
the case of (\ref{3.24}). Details are beyond the scope of this paper.

\section{MASS GENERATION AND RWE THEORY}

In 1974 in the works [13,14] a mechanism of mass generation was proposed
differing from the well-known Higgs mechanism. Later this mechanism has been
identified as a gauge-invariant field mixing. It's essence is as follows.
Two massless systems of equations are considered cooperatively as initial
systems
\begin{subequations}
\label{4.1}
\beqn
\label{4.1a}
\partial _\nu  \psi _{\mu  \nu }   =    0 ,
\\
\label{4.1b}
 -  \partial _\mu  \varphi _\nu     +    \partial _\nu  \varphi
_\mu     +    \psi _{\mu  \nu }     =    0 ,
\eeqn
\end{subequations}
and
\begin{subequations}
\label{4.2}
\beqn
\label{4.2a}
\partial _\mu  \psi _{\mu  \nu  \alpha }     =    0,
\\
\label{4.2b}
 - \partial _\mu \varphi _{\nu \rho } - \partial _\nu \varphi _{\rho \mu } -
\partial _\rho \varphi _{\mu \nu } + \psi _{\mu \nu \rho } = 0,
\eeqn
\end{subequations}
the first system describing an electromagnetic field and the second one
describing
field of Kalb-Ramond. In (\ref{4.2a}) and (\ref{4.2b})
tensor $\psi _{\mu  \nu  \alpha }
$ is considered to be an intensity. Then into the Lagrangian of this system
an additional term is included
\begin{equation}
\label{4.3}
L_{int} = m\varphi _\mu \partial _\nu \varphi _{\mu \nu }
\end{equation}
without violation of the gauge-invariance for the initial Lagrangian $L_0 $.
This term may be formally treated as an interaction of the fields under
study (so-called topological interaction). Varying the Lagrangian $L = L_0 +
L_{int} $ and introducing the pseudovector $\tilde {\psi }_\mu =
\frac{1}{3!}\varepsilon _{\mu \nu \alpha \beta } \psi _{\nu \alpha \beta } $,
we have a system
\begin{subequations}
\label{4.4}
\beqn
\label{4.4a}
\partial _\nu \psi _{\mu \nu } + m\tilde {\psi }_\mu = 0,
\\
\label{4.4b}
 - \partial _\mu \tilde {\psi }_\nu + \partial _\nu \tilde {\psi }_\mu +
m\psi _{\mu \nu } = 0,
\\
\label{4.4c}
\partial _\nu \tilde {\varphi }_{\mu \nu } + \tilde {\psi }_\mu = 0,
\\
\label{4.4d}
 - \partial _\mu \varphi _\nu + \partial _\nu \varphi _\mu + \psi _{\mu \nu
} = 0,
\eeqn
\end{subequations}
where
\begin{equation}
\label{4.5}
\tilde {\varphi }_{\mu \nu } = \frac{1}{2!}\varepsilon _{\mu \nu \alpha
\beta } \varphi _{\alpha \beta } .
\end{equation}
Now in system (\ref{4.4}) we replace $\varphi _\mu $ and $\tilde {\varphi }_{\mu
\nu } $ by the quantities $\tilde {G}_\mu $ and $G_{\mu \nu } $ using the
formulae
\begin{subequations}
\label{4.6}
\beqn
\label{4.6a}
\tilde {G}_\mu = \varphi _\mu - \frac{1}{m}\tilde {\psi }_\mu ,
\\
\label{4.6b}
G_{\mu \nu } = \tilde {\varphi }_{\mu \nu } - \frac{1}{m}\psi _{\mu \nu } .
\eeqn
\end{subequations}
Finally, system (\ref{4.4}) is reduced to the following form:
\begin{subequations}
\label{4.7}
\beqn
\label{4.7a}
\partial _\nu \psi _{\mu \nu } + m\tilde {\psi }_\mu = 0,
\\
\label{4.7b}
 - \partial _\mu \tilde {\psi }_\nu + \partial _\nu \tilde {\psi }_\mu +
m\psi _{\mu \nu } = 0,
\\
\label{4.7c}
\partial _\nu G_{\mu \nu } = 0,
\\
\label{4.7d}
 - \partial _\mu \tilde {G}_\nu + \partial _\nu \tilde {G}_\mu = 0.
\eeqn
\end{subequations}

As seen, system (\ref{4.7}) is reducible with respect to the Lorentz group into
subsystems (\ref{4.7a}), (\ref{4.7b}) and (\ref{4.7c}), (\ref{4.7d}).
The first of them describing a
massive spin 1 particle is interpreted in [13] as an interaction transporter
between open strings. Subsystem (\ref{4.7c}), (\ref{4.7d}) gives no description for a
physical field, as it is associated with zero energy density. However, its
presence is necessary to impart to the latter the status of a
gauge-invariant theory.

Using the formalism of generalized RWE, all the above may be interpreted as
follows. Let us consider a set of representations
\begin{equation}
\label{4.8}
\left( {\frac{1}{2},\frac{1}{2}} \right) \oplus \left(
{\frac{1}{2},\frac{1}{2}} \right)^\prime \oplus 2(1,0) \oplus 2(0,1),
\end{equation}
associated with tensor system (\ref{4.1}), (\ref{4.2}).

It is obvious that on the basis of (\ref{4.8})
one can derive RWE (\ref{eq1.1}) with the
matrices
\begin{equation}
\label{4.9}
\gamma _\mu = \left( {{\begin{array}{*{20}c}
 {\gamma _\mu ^{DK} } \hfill & \hfill \\
 \hfill & {\gamma _\mu ^{DK} } \hfill \\
\end{array} }} \right)  ,
\quad
\gamma _0 = \left( {{\begin{array}{*{20}c}
 {O_4 } \hfill & \hfill & \hfill & \hfill \\
 \hfill & {I{ }_6} \hfill & \hfill & \hfill \\
 \hfill & \hfill & {O_6 } \hfill & \hfill \\
 \hfill & \hfill & \hfill & {I_4 } \hfill \\
\end{array} }} \right)  ,
\end{equation}
where $\gamma _\mu ^{DK} $ are 10-dimensional Duffin-Kemmer matrices.
Introduction into the Lagrangian of a topological term (\ref{4.3}) results in the
changed form of the matrices $\gamma _\mu $ leaving the matrix $\gamma _0
$ unaltered. Substitutions of (\ref{4.6}) are equivalent to the unitary
transformation restoring the form of $\gamma _\mu $ matrix given in (\ref{4.9}).
As this is the case, the matrix $\gamma _0 $ takes the form
\begin{equation}
\label{4.10}
\gamma _0 = \left( {{\begin{array}{*{20}c}
 {mI_{10} } \hfill & \hfill \\
 \hfill & {O_{10} } \hfill \\
\end{array} }} \right).
\end{equation}

In this way we actually arrive at RWE reducible to the ordinary
Duffin-Kemmer equation for a massive spin 1 particle and at the massless
fermionic limit of this equation. Nontrivial nature of the mass generation
method, from the viewpoint of a theory of RWE, consists in the fact that on
passage from the initial massless field(s) to the massive one neither the
form of $\gamma _\mu $ matrices nor the rank of singular $\gamma _0 $ matrix is
affected, the procedure being reduced to permutation of zero and unity
blocks of this matrix only. In the process the number of degrees of freedom
(that is equal to three) for a field system is invariable; it seems as if
the notoph passes its degree of freedom to the photon, that automatically
leads to a massive spin 1 particle.

\section{DISCUSSION AND CONCLUSIONS}

Based on the examples considered, the following important conclusions can be
drawn.

\underline {Conclusion 1.}
\textit{Generalized RWE (\ref{eq1.1}) with the singular matrix }
$\gamma _0 $
\textit{ can describe not only massless but also massive fields (particles).
Featuring the gauge invariance, these equations just form the class
of massive gauge-invariant theories.}

As demonstrated in Sec.~II using equations
(\ref{eq2.34}) and (\ref{eq2.42}) as an example,
a theory of generalized RWE suggests also a variant of the generalized
description for massive and massless fields based on RWE irreducible with
respect to the Lorentz group. Thus, we arrive at the following conclusion.

\underline {Conclusion 2. }
\textit{RWE of the form (\ref{eq1.1}) with the singular matrix }
$\gamma _0 $\textit{ can describe the fields
involving both massive and massless components.
In this case it is more correct to refer to massive-massless
gauge-invariant theories rather than to the massive ones. }

As demonstrated in Sec.~III, within the scope of RWE (\ref{eq1.1}),
on adequate selection of
the Lorentz group representations in a space of the wave function $\psi $
and interpretation of its components, one can give the description of a
massless field not only with helicity $\pm $1 but also with helicity 0 as
well as simultaneous description of the indicated fields. Generalizing this
result for the case of arbitrary spin $S$, we can make the following
conclusion.

\underline {Conclusion 3.} \textit{A theory of the generalized RWE
with the singular matrix} $\gamma _0 $
\textit{makes it possible to describe not only massless fields with maximal
(for the given set of representations) helicity }$\pm S$
\textit{, but also fields with intermediate helicity values
as well as to offer a simultaneous description of these fields. }

It is clear that, all other things being equal, a character of the field
described by equation (\ref{eq1.1}) with the singular matrix $\gamma _0 $ is
dependent on the form of this matrix. To find when the singular matrix
$\gamma _0 $ leads to massless theories and when it results in massive or
massive-massless gauge-invariant theories, we examine the Lorentz structure
of the ``massive'' term $\gamma _0  \psi $ in the foregoing cases. It is
observed that in the case of (\ref{eq2.3}), (\ref{eq2.10}), (\ref{eq2.11})
associated with a massive
gauge-invariant spin 1 theory the matrix $\gamma _0 $ (\ref{eq2.11}) affecting the
wave function $\psi $ (\ref{eq2.3}) in the expression $\gamma _0  \psi $ retains
(without reducing to zero) the Lorentz covariants $\psi _\mu  ,  \psi
_{\mu  \nu } $, on the basis of which an ordinary (of the form (\ref{eq1.2}))
massive spin 1 theory can be framed. But in the case of a massless theory
given by (\ref{eq2.3}), (\ref{3.2}), (\ref{3.3}) the matrix $\gamma _0 $ in the expression
$\gamma _0  \psi $ retains the covariant $   \psi _{\mu  \nu } $, on the
basis of which it is impossible to frame RWE of the form (\ref{eq1.2}) for a massive
particle. A similar pattern is characteristic for the remaining cases: in
all the massive (massive-massless) gauge-invariant theories the matrix
$\gamma _0 $ affecting the wave function $\psi $ retains its covariant
components necessary for framing of an ordinary massive spin 1 or 0 theory;
provided the expression $\gamma _0  \psi $ doesn't involve such a necessary
set of covariants, massless theories can be framed only. This leads us to
the fourth conclusion.

\underline {Conclusion 4. }\textit{Should} \textit{the generalized
RWE (\ref{eq1.1}) with the singular matrix } $\gamma _0 $ in the product
$\gamma _0  \psi $\textit{ retain a set of the Lorentz covariants sufficient
to frame an ordinary (with det }$ \gamma
_0    \ne   0$
\textit{) theory of a massive spin S particle,
this RWE may be associated with a massive gauge-invariant spin S theory.
Otherwise, when this requirement is not fulfilled for any S, RWE (\ref{eq1.1})
can describe a massless field only. }

Proceeding from all the afore-said, we arrive at the following important
though obvious conclusion.

\underline {Conclusion 5.}
\textit{To frame both massive (massive-massless)
gauge-invariant spin $S$ theory and massless theory with intermediate helicity
values from $+S$ to $-S$ we need an extended, in comparison with
 a minimally necessary for the description of this spin (helicity),
 set of the irreducible Lorentz group representations in a space
 of the wave function} $\psi $.

In the present work, when considering spin 1, the above-mentioned extension
has been accomplished by the introduction of scalar representation $\left(
{0 ,  0} \right)$ into a set of the representations given by (\ref{eq2.1}) and of
pseudoscalar representation $\left( {\frac{1}{2} ,  \frac{1}{2}}
\right)^\prime $ -- into a set given by (\ref{eq2.30}). Greater potentialities are
offered by the use of the multiple (recurrent) Lorentz group
representations.

\end{document}